\documentclass[letterpaper, 10 pt, conference]{ieeeconf}  %

\IEEEoverridecommandlockouts                              %

\usepackage{graphics} %
\usepackage{epsfig} %
\usepackage{mathptmx} %
\usepackage{times} %
\usepackage{amsmath} %
\usepackage{amssymb}  %
\usepackage{hyperref}
\usepackage{cleveref}
\usepackage{xcolor}
\usepackage{cite}
\usepackage{newtxtext, newtxmath}
\usepackage[most]{tcolorbox}
\tcbset{breakable}
\usepackage{tikz}
\usetikzlibrary{arrows.meta, positioning, calc}

\newtheorem{remark}{Remark}
\newtheorem{definition}{Definition}

\newcommand{\state}{x}
\newcommand{\control}{u}
\newcommand{\xindex}[2]{\state^{#1}_{#2}}
\newcommand{\uindex}[2]{\control^{#1}_{#2}}
\newcommand{\ufbindex}[2]{\tilde{\control}^{#1}_{#2}}
\newcommand{\xt}{\state_t}
\newcommand{\xtone}{\state_{t+1}}
\newcommand{\uit}{\uindex{i}{t}}

\newcommand{\dyn}{f}
\newcommand{\statedim}{n}
\newcommand{\statedimi}{n^i}
\newcommand{\controldim}{m}
\newcommand{\controldimi}{m^i}
\newcommand{\timeindex}{t}
\newcommand{\statet}{\state_\timeindex}
\newcommand{\playerindex}{i}
\newcommand{\horizon}{T}

\newcommand{\noofplayers}{N}
\newcommand{\policy}{\gamma}
\newcommand{\Cumcost}{J}
\newcommand{\stagecost}{g}
\newcommand{\strategy}{\gamma}
\newcommand{\fbstrat}{\pi}
\newcommand{\valfunc}{V}
\newcommand{\controlvalfunc}{Z}

\newcommand{\ai}{i}
\newcommand{\ami}{-i}

\newcommand{\node}[2]{N^{#1}_{#2}}
\newcommand{\feasiblesetmpn}[2]{F^{#1}_{#2}}
\newcommand{\concatenatedstatempn}{z}
\newcommand{\mpn}{M}

\newcommand{\xnashexample}[2]{\state^{#1, \mathcal{I}}_{#2}}
\newcommand{\unashexample}[2]{\control^{#1, \mathcal{I}}_{#2}}
\newcommand{\ufbnashexample}[2]{\tilde{\control}^{#1, \mathcal{I}}_{#2}}

\newcommand{\LQrunningcost}[3]{g^{#1}_{#2}(\state_{#2},\control^{#1}_{#2},\control^{#3}_{#2})}
\newcommand{\LQcontrolcost}[3]{R^{#1#2}_{#3}}
\newcommand{\LQstatecost}[2]{Q^{#1}_{#2}}

\newcommand{\LQdynstate}[1]{A_{#1}}
\newcommand{\LQdyncontrols}[2]{B^{#1}_{#2}}

\newcommand{\gendynlagplayer}[4]{\eta^{#1}_{#3, (#2,#4)}}
\newcommand{\geninfolagplayer}[4]{\lambda^{#1}_{#3, (#2,#4)}}

\newcommand{\mpncosti}{\mathcal{F}^i}
\newcommand{\mpnfeasibleseti}{\mathcal{C}^i}
\newcommand{\mpndecisionindexseti}{\mathcal{J}^i}
\newcommand{\mpndecisionvariable}{\mathbf{x}}

\newcommand{\mpnedgeset}{E}
\newcommand{\mpndecisiondimensioni}{n^i_\mpndecisionvariable}
\newcommand{\mpnreachable}{R}
\newcommand{\mpndi}{D^i}
\newcommand{\mpndminusi}{D^{-i}}
\newcommand{\mpndvdi}{\mpndecisionvariable^{D^{i}}}
\newcommand{\mpndvdminusi}{\mpndecisionvariable^{D^{-i}}}
\newcommand{\real}{\mathbb{R}}
\newcommand{\solutiongraph}{S}
\newcommand{\solutiongraphi}{S^i}
\DeclareMathOperator*{\argmin}{arg\,min}

\title{\LARGE \bf
Interleaved Information Structures in Dynamic Games: A General Framework with Application to the Linear-Quadratic Case}

\author{Janani S K$^{*1}$, Kushagra Gupta$^{*2}$, Ufuk Topcu$^{3}$, and David Fridovich-Keil$^{3}$%
\thanks{This work was sponsored by the Army Research Laboratory under Cooperative Agreements W911NF-25-2-0021 and ARO W911NF-23-1-0317, the Office of Naval Research under Grant N00014-22-1-2703, the Air Force Office of Scientific Research under Grant FA9550-22-1-0403, and by the National Science Foundation under Grants 2211548 and 2336840. }%
\thanks{* denotes equal contribution}%
\thanks{$^{1}$Janani S K is with Department of Engineering Design, Indian Institute of Technology Madras, Chennai, India
        {\tt\small ed22b017@smail.iitm.ac.in}}%
\thanks{$^{2}$Kushagra Gupta is with the Department of Electrical and Computer Engineering, The University of Texas at Austin,
        Austin, TX 78712, USA
        {\tt\small kushagrag@utexas.edu}}
\thanks{$^{2}$Ufuk Topcu and David Fridovich-Keil are with the Department of Aerospace Engineering and the Oden Institute for Computational Engineering and Sciences, The University of Texas at Austin, Austin, TX 78712, USA {\tt\small utopcu@utexas.edu, dfk@utexas.edu}}%
}

\begin{document}

\maketitle
\thispagestyle{empty}
\pagestyle{empty}

\begin{abstract}
A fundamental problem in noncooperative dynamic game theory is the computation of Nash equilibria under different information structures, which specify the information available to each agent during decision-making. Prior work has extensively studied equilibrium solutions for two canonical information structures: feedback, where agents observe the current state at each time, and open-loop, where agents only observe the initial state. However, these paradigms are often too restrictive to capture realistic settings exhibiting \emph{interleaved} information structures, in which each agent observes only a subset of other agents at every timestep. To date, there is no systematic framework for modeling and solving dynamic games under arbitrary interleaved information structures. To this end, we make two main contributions. First, we introduce a method to model deterministic dynamic games with arbitrary interleaved information structures as Mathematical Program Networks (MPNs), where the network structure encodes the informational dependencies between agents. Second, for linear-quadratic (LQ) dynamic games, we leverage the MPN formulation to develop a systematic procedure for deriving Riccati-like equations that characterize Nash equilibria. Finally, we illustrate our approach through an example involving three agents exhibiting a cyclic information structure.
\end{abstract}

\section{Introduction}\label{section: intro}
Equilibria in non-cooperative dynamic games depend crucially on the underlying \emph{information structure}, which specifies the information available to every agent at each decision-making timestep. Two canonical extremes are typically studied in the literature: open-loop and feedback information structures. Under the feedback information structure, at every timestep, each agent observes the full game state which includes the state of all agents. In contrast, the more restrictive open-loop information assumes that all agents observe only the initial full game state and do not have access to subsequent states when making decisions at later timesteps. Extensive prior work has characterized equilibrium solutions for broad classes of both open-loop and feedback dynamic games \cite{bacsar1998dynamic,fridovich2020efficient,le2022algames,zhu2023sequential,laine2023computation}. However, these existing methods fail to cater to many practical settings in which state information is neither fully available nor completely absent. In realistic multi-agent scenarios, agents often observe the states of only a subset of other agents, leading to \emph{interleaved} information structures where different agents observe different subsets of the system state variables during the game.
\par
Existing works that go beyond the canonical information structures can broadly be categorized into three modeling paradigms. The first allows the dynamic game's information structure to alternate between open-loop and feedback over the time horizon, but still assumes that all agents share identical information at any given time step \cite{zhang2021safe, gupta2024gametheoreticocclusionawaremotionplanning}. The second paradigm studies games with \emph{incomplete} information, where agents lack knowledge of certain game components, such as the costs and dynamics of other agents, and reason by learning these unknown quantities \cite{so2022multimodal, pmlr-v283-soltanian25a}. The third paradigm considers games with \emph{asymmetric} information, in which agents receive private noisy observations of hidden states and reason about other agents’ information by maintaining beliefs based on shared public signals \cite{heydaribeni2019linear, hambly2023linear, swarup2004characterization, schwarting2021stochastic, vasal2021signaling}. A separate but related line of work theoretically characterizes the differences between open-loop and feedback equilibrium solutions in dynamic games \cite{chiu2024extent, gupta2025more}.
\par
However, the above approaches do not readily accommodate interleaved information structures in dynamic games, where each agent may have access to a different subset of the game state at a given time step. Even for the widely studied class of \emph{linear-quadratic} dynamic games---which serve as a fundamental building block for analyzing more complex dynamic games---it remains unclear how to derive Riccati-like equations that yield Nash equilibria for given interleaved information structures. To address this challenge, we build on the recently introduced Mathematical Program Network (MPN) framework \cite{laine2024mathematicalprogramnetworks}, which models collections of interdependent optimization problems as a network and enables the systematic derivation of solutions for such coupled problems. MPNs have previously been used to model interdependencies arising from hierarchical decision-making structures among agents in dynamic games \cite{khan2026efficiently}. In this work, we show that the utility of MPNs for dynamic games extends beyond hierarchical settings; in particular, MPNs provide a natural framework for capturing the interdependencies among agents’ individual optimization problems that arise from interleaved information structures.
\par
\textbf{Contributions:} Motivated by the above discussion, we ask: 
\emph{How can one compute Nash equilibria in linear-quadratic dynamic games where agents possess interleaved information structures, i.e., where each agent observes the states of a (potentially unique) subset of the agents in the game?}
To this end, we make the following contributions:
\begin{enumerate}
    \item We formulate dynamic games with interleaved information structures as MPNs by modeling the time-indexed optimization problems of all agents as a network that captures the interdependencies induced by interleaved information.
    \item Using this formulation, we develop a systematic procedure for deriving \emph{Riccati-like equations} which characterize Nash equilibria for $N$-agent linear-quadratic dynamic games under arbitrary interleaved information structures.
\end{enumerate}

\section{Preliminaries} \label{section: preliminaries}
\subsection{On Mathematical Program Networks (MPNs)}\label{subsection: MPN prelims}
A Mathematical Program Network (MPN) \cite{laine2024mathematicalprogramnetworks} is a directed graph of $K$ decision nodes, each representing a mathematical program. Formally, an MPN is defined by a tuple $(\{\mpncosti, \mpnfeasibleseti, \mpndecisionindexseti\}_{i\in[K]}, \mpnedgeset)$, where $[K]:=\{1,\dots,K\}, K\in\mathbb{Z}^+$. The program corresponding to the $i^{\text{th}}$ decision node has an objective $\mpncosti:\real^{n_\mpndecisionvariable}\rightarrow\real$ over the vector of decision variables $\mpndecisionvariable\in\mpnfeasibleseti\subset\real^{n_\mpndecisionvariable}$. Node $i$ only controls a subset of the entries of $\mpndecisionvariable$, specified by the decision index set $\mpndecisionindexseti \subset [{n_\mpndecisionvariable}]$. The set of directed edges $\mpnedgeset \subset [K]\times[K]$ specifies the network structure and encodes the interdependence between the mathematical programs. An edge $(i,j)\in\mpnedgeset$ indicates that node $j$ is a child of node $i$. Let $\mpndecisiondimensioni = |\mpndecisionindexseti|$, and define the private decision variables of node $i$ as 
$\mpndecisionvariable^i \in \real^{\mpndecisiondimensioni} := [x_j]_{j\in\mpndecisionindexseti}$. We denote the set of reachable node pairs in the MPN by $\mpnreachable \subset [K]\times[K]$, where $(i,j)\in\mpnreachable$ if and only if a path exists from node $i$ to node $j$ obtained by traversing edges in $\mpnedgeset$. The reachable set is useful for identifying the programs whose decisions are dependent on the program at node $i$. In particular, we define $\mpndi := \{i\} \cup \{j : (i,j)\in\mpnreachable\},~\mpndminusi := [K] \setminus \mpndi$. Correspondingly, we partition the decision variables as $\mpndvdi := [\mpndecisionvariable^j]_{j\in\mpndi},~\mpndvdminusi := [\mpndecisionvariable^j]_{j\in\mpndminusi}$. Tracking these interdependencies among the mathematical programs allows us to characterize their solutions through the notion of a \emph{solution graph}. The solution graph $\solutiongraphi$ corresponding to node $i$ yields
\begin{align}\label{eq: solution graph}
\solutiongraphi = 
\left\{
\begin{aligned}
\mpndecisionvariable^*\in&~\real^{n_\mpndecisionvariable}, \textrm{where}~(\mpndecisionvariable^*)^{\mpndi} \textrm{satisfies:}\\
(\mpndecisionvariable^*)^{\mpndi}\in&~\argmin_{\mpndvdi} ~\mpncosti\left(\mpndvdi, (\mpndecisionvariable^*)^{\mpndminusi}\right) \\
&~\textrm{s.t.}~\left(\mpndvdi, (\mpndecisionvariable^*)^{\mpndminusi}\right)\in\mpnfeasibleseti, \\
&\quad~~\,\left(\mpndvdi, (\mpndecisionvariable^*)^{\mpndminusi}\right) \in\solutiongraph^j, (i,j)\in\mpnedgeset.\\
\end{aligned}
\right\}
\end{align}
A solution which is optimal for all the programs in a MPN must belong to the solution graphs of all MPN nodes; such a solution is called an equilibrium.
\begin{definition}[Equilibrium of an MPN] A vector $\mpndecisionvariable^*$ is an equilibrium of an MPN if it is an element of the solution graph of each node. Formally,
    $\mpndecisionvariable^*$ is an equilibrium of an MPN iff $\mpndecisionvariable^*\in\solutiongraph^*$, where $\solutiongraph^*:=\bigcap_{i\in[N]}\solutiongraphi$.
\end{definition}

\subsection{On Dynamic Noncooperative Games}
We consider $N$-agent, discrete-time, deterministic dynamic games with a finite decision-making time horizon of $\horizon$ steps. The state of the game at time $t$ is denoted by $\statet := (\statet^1,\dots,\statet^N),~ \statet^i\in\real^{\statedimi}, ~\sum_i\statedimi=\statedim$, representing the concatenated state of all agents. The state evolves according to the dynamics $\xtone = \dyn_t(\xt, \uindex{1}{t}, \dots, \uindex{N}{t}),~ t \in [\horizon]$ where $\uindex{i}{t} \in \real^{\controldimi}$ denotes the control input of the $i^{\text{th}}$ agent at time $t$. The initial game state $\xindex{}{1}$ is given \emph{a priori}. For brevity, we define $\uindex{-i}{t} := (\uindex{j}{t})_{j\in[N],\, j\neq i}$. For any vector-valued quantity $a$ indexed by agents and/or time, we use $a^{p:q}_{r:s}$ to denote the collection of vectors corresponding to agents $p,p+1,\dots,q$ over times $r,r+1,\dots,s$. Each agent $i$ minimizes a time-additive cost $\Cumcost^i(\xindex{}{1}, \uindex{i}{1:\horizon}, \uindex{-i}{1:\horizon})
= \sum_{t=1}^{\horizon} \stagecost^i_t(\xt, \uit, \uindex{-i}{t})
+ \stagecost^i_{\horizon+1}(\xindex{}{\horizon+1})$. For a linear-quadratic (LQ) dynamic game, the stage cost and dynamics take the form
\begin{align}
\notag\LQrunningcost{i}{t}{-i} = \xt^\top \LQstatecost{i}{t}\xt&+2{q^{i\top}_t}\xt
+ \sum_{j\in[N]} \uindex{j\top}{t} \LQcontrolcost{i}{j}{t}\uindex{j}{t}+2r^{ij\top}_t\uindex{j}{t}, \\
\dyn_t(\xt,\uit,\uindex{-i}{t}) 
&= \LQdynstate{t}\xt + \sum_{j\in[N]} \LQdyncontrols{j}{t}\uindex{j}{t},
\end{align}
where $\LQcontrolcost{i}{i}{t} \succ 0$ and $\LQcontrolcost{i}{j}{t}, \LQstatecost{i}{t} \succeq 0$ for all $i, j$ and $t$. We now review open-loop and feedback Nash equilibria for LQ games, which can be varied as two extreme variants of MPNs for a game.
\subsubsection{Open-Loop Nash Equilibria in Dynamic Games}
Let agent $i$ choose its action at time $t$ using a strategy mapping 
$\strategy^i_t:\real^{\statedim}\rightarrow\real^{\controldimi}$. Under the open-loop information structure, each agent observes only the initial game state and does not have access to subsequent game states. Consequently, the control actions of each agent are determined solely based on the initial state,i.e., 
$\uindex{i}{t}=\strategy^i_t(\xindex{}{1})$. An open-loop Nash equilibrium for an LQ dynamic game consists of control sequences $\uindex{i,\text{OL}}{1:\horizon}$ and the corresponding state trajectory $\xindex{\text{OL}}{1:\horizon+1}$ (with $\xindex{\text{OL}}{1}=\xindex{}{1}$) such that for all agents $i\in[\noofplayers]$,
\[
\Cumcost^i(\xindex{}{1},\uindex{i,\text{OL}}{1:\horizon},\uindex{-i,\text{OL}}{1:\horizon})
\le
\Cumcost^i(\xindex{}{1},\tilde{\control}^i_{1:\horizon},\uindex{-i,\text{OL}}{1:\horizon})
\quad
\forall~\tilde{\control}^i_{1:\horizon}\in\real^{\controldimi\times\horizon}.
\]

Computing an open-loop Nash equilibrium in an $\noofplayers$-agent LQ dynamic game amounts to solving $\noofplayers$ optimization problems. For agent $i$, the problem is
\begin{align}\label{eq: ol opt problem}
\begin{aligned}
\uindex{i,\text{OL}}{1:\horizon} \in 
&\argmin_{\xindex{}{2:\horizon+1},\,\uindex{i}{1:\horizon}}
\Cumcost^i\left(\xindex{}{1},\uindex{i}{1:\horizon},\uindex{-i,\text{OL}}{1:\horizon}\right), \\
&\textrm{s.t.} \quad 
\xtone=\dyn_t(\xt,\uindex{i}{t},\uindex{-i,\text{OL}}{t}),
\qquad t\in[\horizon].
\end{aligned}
\end{align}

\subsubsection{Feedback Nash Equilibria in Dynamic Games}

Under the feedback information structure, all agents observe the current game state $\xt$ when making decisions at time $t$. Accordingly, a feedback Nash equilibrium is defined in terms of feedback strategy mappings 
$\fbstrat^i_t:\real^{\statedim}\rightarrow\real^{\controldimi}$ for $t\in[\horizon]$, state value functions 
$\valfunc^i_t:\real^{\statedim}\rightarrow\real$ for $t\in[\horizon+1]$, and state-control value functions 
$\controlvalfunc^i_t:\real^{\statedim}\times\prod_{i\in[\noofplayers]}\real^{\controldimi}\rightarrow\real$ for $t\in[\horizon]$, which satisfy the following backward recursion for each $i\in[\noofplayers]$:
\begin{align} \label{eq: fb val func recursions}
\begin{aligned}
\valfunc^i_{\horizon+1}(\xindex{}{\horizon+1})
&:= \stagecost^i_{\horizon+1}(\xindex{}{\horizon+1}), \\
\controlvalfunc^i_t(\xt,\uindex{i}{t},\uindex{-i}{t})
&:= \stagecost^i_t(\xt,\uindex{1:\noofplayers}{t})
+ \valfunc^i_{t+1}\!\left(\dyn(\xt,\uindex{i}{t},\uindex{-i}{t})\right), \\
\valfunc^i_t(\xt)
&:= \controlvalfunc^i_t(\xt,\fbstrat^{i}_{t}(\xt),\fbstrat^{-i}_{t}(\xt)),
\qquad \forall~t\in[\horizon].
\end{aligned}
\end{align}

Let $\uindex{i,\text{FB}}{1:\horizon}$ and $\xindex{\text{FB}}{1:\horizon+1}$ denote the control and state trajectories corresponding to the feedback Nash equilibrium for all agents $i$, with $\xindex{\text{FB}}{1}=\xindex{}{1}$. The feedback strategies $\fbstrat^i_t$ are defined so that for each time $t\in[\horizon]$,
\begin{align}
\controlvalfunc^i_t(\xt,\uindex{i,\text{FB}}{t},\uindex{-i,\text{FB}}{t})
\le
\controlvalfunc^i_t(\xt,\tilde{\control}^i_t,\uindex{-i,\text{FB}}{t}),~ \forall ~\tilde{\control}^i_t\in\real^{\controldimi}.
\end{align}

Computing a feedback Nash equilibrium amounts to solving a sequence of nested equilibrium problems, one at each time step coupling all the agents' decisions. For agent $i$ at time $t\in[\horizon]$, the corresponding optimization problem is \cite{laine2023computation}:
\begin{align}\label{eq: fb opt problem}
\begin{aligned}
\min_{\xindex{}{s:\horizon+1},\,\uindex{i}{t:\horizon},\,\ufbindex{-i,\text{FB}}{t+1:\horizon}}
\sum_{s=t}^{\horizon}
\stagecost_s(\xindex{}{s},\uindex{i}{s},&\ufbindex{-i,\text{FB}}{s})
+ \stagecost_{\horizon+1}(\xindex{}{\horizon+1}) \\
\textrm{s.t.}\quad
\xindex{}{s+1} - \dyn_s(\xindex{}{s},\uindex{i}{s},\ufbindex{-i,\text{FB}}{s}) &= 0,
\qquad t\leq s\leq \horizon, \\
\ufbindex{-i,\text{FB}}{s} - \fbstrat^{-i}_s(\xindex{}{s}) &= 0,
\qquad t+1\leq s\leq \horizon .
\end{aligned}
\end{align}
\begin{remark}\label{remark: fb interdependence}
    The final set of feedback constraints in \Cref{eq: fb opt problem} are an example of the game's information structure inducing interdependencies between the agents' optimization problems, which will be modeled as edges in accordance with \Cref{subsection: MPN prelims}. This interdependence enforces strong Bellman consistency required to satisfy \Cref{eq: fb val func recursions}.  
\end{remark}

\section{Formulating Interleaved Information Games as MPNs} \label{section: mpn formulation}
We now present our main contributions. As a warm-up, we first show how two-agent open-loop and feedback dynamic games can be formulated as MPNs. The insights gained from these canonical cases will guide the systematic construction of MPNs for dynamic games with arbitrarily interleaved information structures. We refer to the two agents in both open-loop and feedback games as $\ai$ and $\ami$, respectively.

\subsubsection{MPNs for open-loop games}
Consider the cost-to-go that agent $\ai$ minimizes at the decision-making time step $\timeindex\in[\horizon]$, $\Cumcost^\ai_\timeindex(\xindex{}{\timeindex:\horizon+1}, \uindex{\ai}{\timeindex:\horizon}, \uindex{\ami}{\timeindex:\horizon}):=\sum_{s=\timeindex}^\horizon \stagecost_s^\ai(\xindex{}{s}, \uindex{\ai}{s}, \uindex{\ami}{s}) + \stagecost_{\horizon+1}^\ai(\xindex{}{\horizon+1})$. From \Cref{eq: ol opt problem}, two observations follow. First, the optimization problem of agent $\ai$ at any timestep $\timeindex$ influences the actions that $\ai$ selects at future timesteps. Second, because agent $\ami$ must choose an open-loop strategy, its action at any time after $t$ cannot depend on agent $i$'s actions at time $t$. Therefore, for an open-loop dynamic game with a decision-making horizon of $\horizon$ steps, we construct an MPN with $\horizon$ nodes per agent. The $\timeindex^{\text{th}}$ node corresponding to agent $\ai$, denoted by $\node{\ai}{\timeindex}$, represents the problem of minimizing the cost-to-go $\Cumcost^\ai_\timeindex$.

The interdependence across time induces edges between consecutive nodes of the same agent. Let $\mathcal{A}:=\{\ai,\ami\}$, and let the concatenated decision variables for agent $j$ across all nodes be denoted by $\concatenatedstatempn^j$, for $j\in\mathcal{A}$. For brevity, we denote the decision variables at node $\node{j}{t}$ by $\concatenatedstatempn^j_t$, consisting of $\{\xindex{}{t+1:\horizon+1}, \uindex{j}{t:\horizon}\}$. The variables $\concatenatedstatempn^j_t$ are constrained to lie in the dynamically feasible subset $\feasiblesetmpn{j}{t}\subset(\real^\statedim\times\real^{\controldim^j})^{\horizon-t+1}$. The edge set for agent $j$ is $\mpnedgeset^{j, OL}$, consisting of $\horizon-1$ edges connecting $\node{j}{t}$ to $\node{j}{t+1}$ for all $\timeindex\in[\horizon-1]$. Let $\mpnedgeset^{OL}=\{\mpnedgeset^{\ai, OL},\mpnedgeset^{\ami, OL}\}$. Then the open-loop MPN $\mpn^{\text{OL}}$, shown in \Cref{fig: canonical mpns}, is
\begin{align}\notag
    \mpn^{\text{OL}} = \left\{\left\{\{\Cumcost^{j}_{s}\}_{s\in[\horizon]}, \{\feasiblesetmpn{j}{s}\}_{s\in[\horizon]}, \{\concatenatedstatempn^j_s\}_{s\in[\horizon]}\right\}_{j\in\mathcal{A}}, \mpnedgeset^{OL}\right\}.
\end{align}

\subsubsection{MPNs for feedback games}
\Cref{remark: fb interdependence} shows that in the feedback setting, at any timestep an agent's optimization problem depends not only on its own future decisions but also on the future strategies of other agents. Consequently, the optimization problems across agents become coupled. As in the open-loop case, we construct an MPN with $\horizon$ nodes per agent, where node $\node{j}{t}$ represents agent $j$'s optimization problem at time $t$. The node set therefore is unchanged. However, the interdependence induced by feedback information requires augmenting both the decision variables and the edge set. In particular, each agent $j\in\mathcal{A}$ introduces additional variables representing its reasoning about the other agent $-j$'s future decisions made at $\node{-j}{\timeindex+1}$. Thus, for $j\in\mathcal{A}$, $\concatenatedstatempn^{j,\text{FB}}_{t}=\concatenatedstatempn^{j,\text{OL}}_{t}\bigcup\{\ufbindex{-j, \text{FB}}{\timeindex+1:\horizon}\}$, $\feasiblesetmpn{j, \text{FB}}{t}\subset(\real^\statedim\times\real^{\controldim^j})^{T-t+1}\times\real^{\controldim^{-j}})^{T-t}$, $\mpnedgeset^{j,\text{FB}}=\mpnedgeset^{j, \text{OL}}\bigcup\{\node{j}{t}\rightarrow\node{-j}{\timeindex+1}\}$, and $\mpnedgeset^{\text{FB}}=\{\mpnedgeset^{i, \text{FB}}, \mpnedgeset^{-i, \text{FB}}\}$. Then the feedback MPN, shown in \Cref{fig: canonical mpns}, is 
\begin{align}\notag
        \mpn^{\text{FB}} = \left\{\left\{\{\Cumcost^{j}_{s}\}_{s\in[\horizon]}, \{\feasiblesetmpn{j, FB}{s}\}_{s\in[\horizon]}, \{\concatenatedstatempn^{j, FB}_s\}_{s\in[\horizon]}\right\}_{j\in\mathcal{A}}, \mpnedgeset^{FB}\right\}.
\end{align}
\subsection{MPNs for arbitrarily interleaved information}
We are now ready to construct MPNs for dynamic games with arbitrary interleaved information structures. We assume that this information structure is known to all agents at the starting of the game. Consider an $\noofplayers$-agent game and any pair of agents $i$ and $j$. The information relationship between these agents must fall into one of four cases at every decision-making timestep time $\timeindex$: (i) mutual feedback, where both agents observe each other; (ii) mutually open-loop, where neither agent observes the other; (iii) one-sided observation, where $i$ observes $j$ but not vice versa; or (iv) one-sided observation in the opposite direction, where $j$ observes $i$ but not vice versa. 

The open-loop and feedback MPN constructions presented earlier already cover the first two cases and provide insight into the remaining asymmetric cases. Without loss of generality, we therefore focus on the one-sided observation case, $\mathcal{I}$, in which $i$ observes $j$ but not vice versa. At decision-making timestep $\timeindex$, agent $i$ incorporates the state(s) of agent $j$ into its decision-making process. Consequently, agent $j$'s optimization problem at $\timeindex$ influences agent $i$'s optimization problem at $\timeindex+1$. In contrast, since agent $j$ does not observe agent $i$, the optimization problem of agent $j$ at $\timeindex$ does not depend on agent $i$'s past decisions.  \emph{Thus, in addition to the agent-wise temporal interdependencies, the MPN includes edges $\node{j}{\timeindex}\rightarrow\node{i}{\timeindex+1}$ for all timesteps $\timeindex$ in which agent $i$ observes agent $j$.}

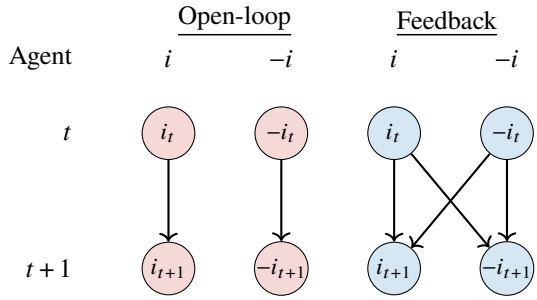
\begin{figure}[h]
\centering

\def\Agents{2}
\def\Horizon{2}

\def\NodeRadius{0.35}
\def\XSep{1.5}
\def\YSep{1.8}

\def\MPNShift{3}

\pgfmathtruncatemacro{\Hminusone}{\Horizon-1}

\definecolor{agent1}{RGB}{203,67,53}
\definecolor{agent2}{RGB}{203,67,53}

\begin{tikzpicture}[
node_style/.style={
circle,
draw,
minimum size=2*\NodeRadius cm,
inner sep=0pt,
font=\small
},
dynedge/.style={->, thick},
infoedge/.style={->, thick}
]

\begin{scope}

\foreach \t in {1,...,\Horizon}{

    \pgfmathtruncatemacro{\k}{\t-1}
    \ifnum\k=0
        \def\timelabel{t}
    \else
        \edef\timelabel{t+\k}
    \fi

    \foreach \a in {1,...,\Agents}{

        \ifcase\a
        \or \def\col{agent1}\def\agentlabel{\ai}
        \or \def\col{agent2}\def\agentlabel{\ami}
        \fi

        \node[node_style, fill=\col!20]
        (L\a\t)
        at (\a*\XSep, -\t*\YSep)
        {$\agentlabel_{\timelabel}$};

    }

    \node[left=0.8cm of L1\t] {$\timelabel$};
}

\foreach \a in {1,...,\Agents}{

\ifcase\a
\or \def\agentlabel{i}
\or \def\agentlabel{-i}
\fi

\node[above=0.4cm of L\a1] {$\agentlabel$};

}

\node[above=0.8cm of L11] {\quad\quad\quad\quad\quad \underline{Open-loop}};

\node[left=0.8cm of L11, yshift=1.0cm] {Agent};

\foreach \a in {1,...,\Agents}{
\foreach \t in {1,...,\Hminusone}{
\pgfmathtruncatemacro{\tnext}{\t+1}
\draw[dynedge] (L\a\t) -- (L\a\tnext);
}
}

\end{scope}

\definecolor{agent1}{RGB}{46,134,193}
\definecolor{agent2}{RGB}{46,134,193}

\begin{scope}[xshift=\MPNShift cm]

\foreach \t in {1,...,\Horizon}{

    \pgfmathtruncatemacro{\k}{\t-1}
    \ifnum\k=0
        \def\timelabel{t}
    \else
        \edef\timelabel{t+\k}
    \fi

    \foreach \a in {1,...,\Agents}{

        \ifcase\a
        \or \def\col{agent1}\def\agentlabel{\ai}
        \or \def\col{agent2}\def\agentlabel{\ami}
        \fi

        \node[node_style, fill=\col!20]
        (R\a\t)
        at (\a*\XSep, -\t*\YSep)
        {$\agentlabel_{\timelabel}$};

    }
}

\foreach \a in {1,...,\Agents}{

\ifcase\a
\or \def\agentlabel{i}
\or \def\agentlabel{-i}
\fi

\node[above=0.4cm of R\a1] {$\agentlabel$};

}

\node[above=0.8cm of R11] {\quad\quad\quad\quad \underline{Feedback}};

\foreach \a in {1,...,\Agents}{
\foreach \t in {1,...,\Hminusone}{
\pgfmathtruncatemacro{\tnext}{\t+1}
\draw[dynedge] (R\a\t) -- (R\a\tnext);
}
}

\draw[infoedge] (R11) -- (R22);
\draw[infoedge] (R21) -- (R12);

\end{scope}

\end{tikzpicture}

\caption{MPN building blocks across successive time steps for dynamic games with canonical information structures.}
\label{fig: canonical mpns}

\end{figure}

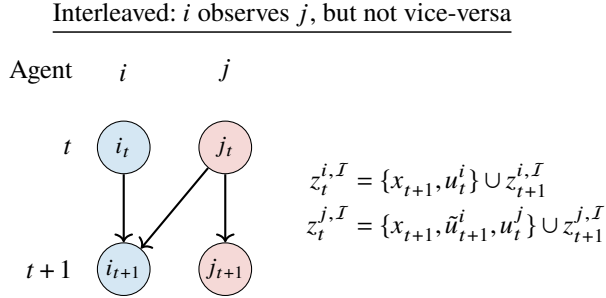
\begin{figure}[h]
\centering

\def\Agents{2}
\def\Horizon{2}

\def\NodeRadius{0.35}
\def\XSep{1.5}
\def\YSep{1.8}

\def\MPNShift{3}

\pgfmathtruncatemacro{\Hminusone}{\Horizon-1}

\definecolor{agent1}{RGB}{46,134,193}
\definecolor{agent2}{RGB}{203,67,53}

\begin{center}
\underline{Interleaved: $i$ observes $j$, but not vice-versa}
\end{center}

\begin{minipage}[t]{0.52\linewidth}
\centering

\begin{tikzpicture}[scale=0.9,
node_style/.style={
circle,
draw,
minimum size=2*\NodeRadius cm,
inner sep=0pt,
font=\small
},
dynedge/.style={->, thick},
infoedge/.style={->, thick}
]

\begin{scope}

\foreach \t in {1,...,\Horizon}{

    \pgfmathtruncatemacro{\k}{\t-1}
    \ifnum\k=0
        \def\timelabel{t}
    \else
        \edef\timelabel{t+\k}
    \fi

    \foreach \a in {1,...,\Agents}{

        \ifcase\a
        \or \def\col{agent1}\def\agentlabel{i}
        \or \def\col{agent2}\def\agentlabel{j}
        \fi

        \node[node_style, fill=\col!20]
        (L\a\t)
        at (\a*\XSep, -\t*\YSep)
        {$\agentlabel_{\timelabel}$};

    }

    \node[left=0.2cm of L1\t] {$\timelabel$};
}

\foreach \a in {1,...,\Agents}{

\ifcase\a
\or \def\agentlabel{i}
\or \def\agentlabel{j}
\fi

\node[above=0.4cm of L\a1] {$\agentlabel$};
}

\node[left=0.2cm of L11, yshift=1.0cm] {Agent};

\foreach \a in {1,...,\Agents}{
\foreach \t in {1,...,\Hminusone}{
\pgfmathtruncatemacro{\tnext}{\t+1}
\draw[dynedge] (L\a\t) -- (L\a\tnext);
}
}

\draw[infoedge] (L21) -- (L12);

\end{scope}

\end{tikzpicture}
\end{minipage}
\hfill
\begin{minipage}[t]{0.40\linewidth}
\vspace{-2.1cm}
\begin{align}
\notag
\concatenatedstatempn^{i,\mathcal{I}}_t
&=
\{\xindex{}{t+1},\uindex{i}{t}\}
\cup
\concatenatedstatempn^{i,\mathcal{I}}_{t+1}
\\
\notag
\concatenatedstatempn^{j,\mathcal{I}}_t
&=
\{\xindex{}{t+1},\ufbindex{i}{t+1},\uindex{j}{t}\}
\cup
\concatenatedstatempn^{j,\mathcal{I}}_{t+1}
\end{align}
\end{minipage}

\caption{MPN building block across successive timesteps for a dynamic game in which agents $i$ and $j$ exhibit a non-canonical interleaved information structure. Additional edges may appear depending on the information relationships with other agents in the game.}
\label{fig:interleaved_mpn}

\end{figure}

This further implies that MPN decision variables for agent $j$ at time $\timeindex$, $\concatenatedstatempn^{j, \mathcal{I}}_\timeindex$ should be augmented to include decision variables of $i$ at its future timestep node $\node{i}{\timeindex+1}$. These arguments are precisely presented in \Cref{fig:interleaved_mpn}, and are simply reversed for case (iv). To construct the MPN for the entire game, the above argument is applied to every pair of agents, at every timestep. Notably, the MPN construction accommodates interleaved information in which an agent $i$ observes different subsets of agents across timesteps. Successive timesteps in any arbitrary interleaved information structure can be decomposed into one of the four canonical cases for which we have constructed MPNs, as illustrated in \Cref{fig: canonical mpns,fig:interleaved_mpn}.

\section{Finding Nash Equilibria in Interleaved Information Games}\label{section: deriving riccati}
We now present our second contribution: a systematic method for computing Nash equilibria in dynamic games with arbitrary interleaved information structures, leveraging the MPN construction procedure developed in \Cref{section: mpn formulation}. Given an MPN representation of a game, a solution graph is constructed for each node using \Cref{eq: solution graph}. Under appropriate constraint qualifications, the optimal decision variables at each node satisfy the first-order necessary optimality conditions—the \emph{Karush-Kuhn-Tucker (KKT) conditions} \cite{nocedal2006numerical} associated with \Cref{eq: solution graph}.
In the special case of linear-quadratic (LQ) games, the affine structure of the constraints and convexity of agent cost functions ensure that these constraint qualifications hold. \emph{Further, for LQ games, due to convexity, the KKT conditions are not only necessary but also sufficient for optimality, and reduce to a system of Riccati-like equations that characterize the Nash equilibrium.}

It follows that the Nash equilibrium of an LQ game with an interleaved information structure can be computed via the following procedure: (i) formulate the game as an MPN, (ii) construct the solution graph at each node, and (iii) solve the resulting system of Riccati-like equations obtained by concatenating the KKT conditions across all nodes. We now formalize this procedure for an LQ game with any possible interleaved information structure.
\begin{tcolorbox}[title=A Systematic Procedure for Deriving Riccati-like Equations in LQ Games with Interleaved Information, colback=white, colframe=black, breakable]
Given an $\noofplayers$-agent LQ dynamic game with interleaved information, and decision-making time horizon $\horizon$: 
\begin{enumerate}
    \item Construct the corresponding MPN as follows:
\begin{enumerate}
    \item \textbf{Nodes.} For each agent $i\in[\noofplayers]$ and timestep $t\in[\horizon]$, create a node $\node{i}{t}$ representing agent $i$'s problem of optimizing $\Cumcost^i_t$ at time $t$.
    \item \textbf{Dynamical Constraints.} Add the dynamical constraints for times $\{t,t+1,\dots,\horizon\}$ to the constraint set $\mathcal{C}^i_t$ corresponding to $\node{i}{t}$.
    \item 
    \textbf{Temporal edges.} For every agent $i$ and timestep $t\in[\horizon-1]$, include an edge $\node{i}{t}\rightarrow\node{i}{t+1}$ capturing the dependence of agent $i$'s actions at time $t$ on its future actions.
    \item \textbf{Observation edges.} For any pair of agents $(i,j)$ and timestep $t$, if agent $i$ observes agent $j$ at time $t$, include an edge $\node{j}{t}\rightarrow\node{i}{t+1}$ capturing the dependence of agent $i$'s future decision problem on agent $j$'s current decision at time $t$.
\end{enumerate}
This MPN encodes the interdependencies between the agents' optimization problems induced by the interleaved information structure. 
\item Create the solution graph for all nodes of the constructed MPN, according to \Cref{eq: solution graph}. For each solution graph, write the corresponding KKT conditions.
\item Concatenating the KKT conditions of all nodes yields Riccati-like equations. Jointly solving them yields the Nash equilibrium of the game.
\end{enumerate} 
\end{tcolorbox}

\section{Illustrative Example}
We now present an example illustrating our contributions, following the procedure outlined in \Cref{section: deriving riccati}. Specifically, we consider a three-player LQ game with a circular information structure.
\begin{figure}[h]
\centering

\def\Agents{3}
\def\Horizon{3}

\def\NodeRadius{0.35}
\def\XSep{2}
\def\YSep{1.8}

\pgfmathtruncatemacro{\Hminusone}{\Horizon-1}

\definecolor{agent1}{RGB}{203,67,53}
\definecolor{agent2}{RGB}{46,134,193}
\definecolor{agent3}{RGB}{39,174,96}

\begin{center}
\underline{Cyclical Interleaved Information Structure}
\end{center}

\begin{tikzpicture}[scale=0.9,
node_style/.style={
circle,
draw,
minimum size=2*\NodeRadius cm,
inner sep=0pt,
font=\small,
path picture={
    \fill[agent1!30]
        (path picture bounding box.south west)
        rectangle
        (path picture bounding box.north);
    \fill[agent2!30]
        (path picture bounding box.south)
        rectangle
        (path picture bounding box.north east);
}
},
dynedge/.style={->, thick},
infoedge/.style={->, thick}
]

\begin{scope}

\foreach \t in {1,...,\Horizon}{

    \pgfmathtruncatemacro{\k}{\t}
    \def\timelabel{\k}

    \foreach \a in {1,...,\Agents}{

        \ifcase\a
        \or \def\agentlabel{1}
        \or \def\agentlabel{2}
        \or \def\agentlabel{3}
        \fi

        \node[node_style]
        (L\a\t)
        at (\a*\XSep, -\t*\YSep)
        {$\agentlabel_{\timelabel}$};

    }

    \node[left=0.2cm of L1\t] {$t=\timelabel$};
}

\foreach \a in {1,...,\Agents}{

\ifcase\a
\or \def\agentlabel{1}
\or \def\agentlabel{2}
\or \def\agentlabel{3}
\fi

\node[above=0.4cm of L\a1] {$\agentlabel$};
}

\node[left=0.2cm of L11, yshift=1.0cm] {Agent};

\foreach \a in {1,...,\Agents}{
\foreach \t in {1,...,\Hminusone}{
\pgfmathtruncatemacro{\tnext}{\t+1}
\draw[dynedge] (L\a\t) -- (L\a\tnext);
}
}

\draw[infoedge] (L21) -- (L12);
\draw[infoedge] (L22) -- (L13);
\draw[infoedge] (L12) -- (L13);
\draw[infoedge] (L22) -- (L23);
\draw[infoedge] (L32) -- (L33);
\draw[infoedge] (L31) -- (L22);
\draw[infoedge] (L32) -- (L23);
\draw[infoedge] (L11) -- (L32);
\draw[infoedge] (L12) -- (L33);

\end{scope}

\end{tikzpicture}

\caption{MPN for a three-agent, three timestep game with a cyclical information structure: agent 1 observes agent 2, who in turn observes agent 3, who in turn observes agent 1.}
\label{fig:cyclic_example_mpn}

\end{figure}
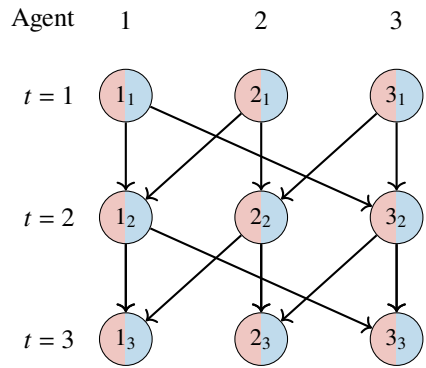

Consider an LQ game with three agents---$\{1,2,3\}$, and a decision-making time horizon of $\horizon=3$ steps. We assume the following circular interleaved information structure:
\begin{itemize}
    \item Agent 1 observes agent 2, but not vice-versa.
    \item Agent 2 observes agent 3, but not vice-versa.
    \item Agent 3 observes agent 1, but not vice-versa.
\end{itemize}
For a positive semi-definite matrix $M$ and a vector $v$, we denote $\|v\|_M=\sqrt{v^\top M v}$ and $\|v\|=\sqrt{v^\top v}$. For $i\in[3]$, we assume that the $i^{\text{th}}$ agent has quadratic stagewise and terminal costs, $\stagecost^i_t, t\in[3]$ and $\stagecost^i_4$, respectively. We have
\begin{align*}
    \stagecost^i_t(u_t,x_t)=\|&x^i_t-x^i_g\|^2+ \|u^i_t\|^2_{\LQcontrolcost{i}{i}{t}}+\sum_{j\neq i} \|x^i_t-x^j_t-p^{ij}_t\|^2_{\LQstatecost{ij}{t}}\\
   \stagecost^i_4(u_t,x_t)=&\|x^i_4-x^i_g\|^2+\sum_{j\neq i} \|x^i_4-x^j_4-p^{ij}_4\|^2_{\LQstatecost{ij}{4}},
\end{align*}
where $\LQcontrolcost{i}{i}{t}\succ0~\forall ~t\in[3]$, $\LQstatecost{ij}{t}\succeq0~\forall~ t\in[4]$, $x^i_g$ represents the goal state for agent $i$, and $p^{ij}_t$ represents the desired state difference between agents $i$ and $j$ at time $t$. We assume that agent $i$'s state evolves only due its own control, and that the linear dynamics for agent $i$ are given by
\begin{align*}
    x^i_{t+1}=f^i_t(x^i_t, u^i_t) := A^i_tx^i_t+B^i_tu^i_t, ~t\in[3].
\end{align*}
The MPN for this cyclic information game, presented in \Cref{fig:cyclic_example_mpn}, is made according to the procedure outlined in \Cref{section: deriving riccati}. Let the state and action of agent $i$ corresponding to the Nash equilibrium of the game at time $t$ be $\xnashexample{i}{t}$ and $\unashexample{i}{t}$ respectively. Further, let the state accessible to agent $i$ at time $t$ be $X^i_t$. Then, we denote the Nash equilibrium decision for agent $i$ at time $t$ as $\unashexample{i}{t}=\policy^i_t(X^i_t)$. As an example of constructing the solution graph at a node, consider agent 1 at time $t=1$. Agent 1 has the decision variables $\concatenatedstatempn^1_1=\{x_{2:4}, \uindex{1}{1:3}, \ufbindex{2}{3}, \ufbindex{3}{2:3}\}$ at node $\node{1}{1}$. For $i=\{2,3\}, t\in[3]$, let $\uindex{i\phi^1}{t}=\ufbindex{i}{t}$ if $\ufbindex{i}{t}\in\concatenatedstatempn^1_{1}$, else $\uindex{i\phi^1}{t}=\unashexample{i}{t}$. From the MPN given in \Cref{fig:cyclic_example_mpn}, we can construct the solution graph $\solutiongraph^1_1$ for node $\node{1}{1}$ through \Cref{eq: solution graph}, yielding the optimization problem
\begin{align*}
        \xindex{\mathcal{I}}{2:4}, \unashexample{1}{1:3}, \ufbnashexample{2}{3}, &\ufbnashexample{3}{2:3} \in \argmin_{\xindex{}{2:4}, \uindex{1}{1:3}, \ufbindex{2}{3}, \ufbindex{3}{2:3}}\Cumcost^1_1(x_{1:4}, \uindex{1}{1:3})\\
    \textrm{s.t.}~&\left\{\begin{aligned}
        &\xtone-\dyn_t(\xt, \uindex{1}{t}, \uindex{2\phi^1}{t}, \uindex{3\phi^1}{t})=0,~ t\in[3]\\
    &\ufbindex{3}{2:3} \in \solutiongraph^3_2 ~\textrm{s.t}~\ufbindex{3}{3}\in\solutiongraph^3_3\\
    &\ufbindex{2}{3}\in\solutiongraph^2_3\\
    &\uindex{1}{2:3}\in\solutiongraph^1_2~\textrm{s.t.}~\uindex{1}{3}\in\solutiongraph^1_3.
    \end{aligned}\right.
\end{align*}
The optimization problems at other nodes can be transcribed similarly. We now proceed to derive the KKT conditions for node $\node{1}{1}$. Observe that three distinct types of constraints exist. For the problem being considered at node $\node{i}{t}$, we denote the Lagrange multipliers for each constraint type as follows:
\begin{itemize}
    \item $\gendynlagplayer{\playerindex}{j}{\timeindex}{k}$ is the multiplier associated with agent $j$'s state dynamics being feasible at time $k\geq t$.
    \item $\geninfolagplayer{i}{i}{t}{k}$ is the multiplier associated with the dependence of agent $i$'s action at $t$ on its future action taken at time $k$.
    \item $\geninfolagplayer{i}{j}{t}{k}$ is the multiplier associated with the dependence of agent $i$'s action at $t$ on agent $j$'s future action taken at time $k$, where this dependence occurs due to the interleaved information structure.
\end{itemize}
Note that the first two types of multipliers correspond to constraints that remain the same for any interleaved information structure, while the last type of  constraints strongly depend on the information structure.
The Lagrangians for agent 1 at times $1,2$, and $3$ thus become
\begin{align}
    \notag\mathcal{L}^1_3=\Cumcost^1_3+&{(\eta^{1}_{3, (1,4)})}^\top(x^1_4-A^1_3x^1_3-B^1_3u^{1}_3)\\
    +&\sum_{j=2,3}(\eta^1_{3, (j,4)})^\top(x^j_4-A^j_3x^j_3-B^j_3\unashexample{j}{3}), \label{eq: l13}\\
    \notag\mathcal{L}^1_2=\Cumcost^1_2+&(\eta^1_{2,(1,3)})^\top(x^1_3-A^1_2x^1_2-B^1_2u^{1}_2)\\
    \notag+&\sum_{j=2,3}(\eta^1_{2, (j,3)})^\top(x^j_3-A^j_2x^j_2-B^j_2\unashexample{j}{2})\\
    \notag+&(\eta^1_{2,(2,4)})^\top(x^2_4-A^2_3x^2_3-B^2_3\unashexample{2}{3})\\
    \notag+&(\eta^1_{2,(1,4)})^\top(x^1_4-A^1_3x^1_3-B^1_3u^{1}_3)\\
    \notag+&(\eta^1_{2,(3,4)})^\top(x^3_4-A^3_3x^3_3-B^3_3\ufbindex{3}{3})\\
    \notag+&(\lambda^1_{2,(3,3)})^\top\left(\ufbindex{3}{3}-\policy^{3}_3(x^1_3,x^3_3)\right)\\
    +&(\lambda^1_{2,(1,3)})^\top\left(u^1_3-\policy^{1}_3(x^1_3,x^2_3)\right) \label{eq: l12}\\
    \notag\mathcal{L}^1_1 = \Cumcost^1_1 + & (\eta^1_{1,(1,4)})^\top(x^1_4-A^1_3x^1_3-B^1_3u^{1}_3)\\
    \notag+ & \sum_{j=2,3} (\eta^1_{1,(j,4)})^\top(x^j_4-A^j_3x^j_3-B^j_3\ufbindex{j}{3})\\ %
    \notag+&(\eta^1_{1,(2,3)})^\top(x^2_3-A^2_2x^2_2-B^2_2\unashexample{2}{2})\\
    \notag+&(\eta^1_{1,(1,3)})^\top(x^1_3-A^1_2x^1_2-B^1_2u^{1}_2)\\
    \notag+&(\eta^1_{1,(3,3)})^\top(x^3_3-A^3_2x^3_2-B^3_2\ufbindex{3}{2})\\
    \notag+&(\eta^1_{1,(1,2)})^\top(x^1_2-A^1_1x^1_1-B^1_1u^{1}_1)\\
    \notag+&\sum_{j=2,3}(\eta^1_{1,(j,2)})^\top(x^j_2-A^j_1x^j_1-B^j_1\unashexample{j}{1})\\
    \notag+&(\lambda^1_{1,(3,3)})^\top\left(\ufbindex{3}{1}-\policy^{3}_3(x^1_3,x^3_3)\right)\\
    \notag+&(\lambda^1_{1,(3,2)})^\top\left(\ufbindex{3}{2}-\policy^{3}_2(x^1_2,x^3_2)\right)\\
    \notag+&(\lambda^1_{1,(2,3)})^\top\left(\ufbindex{2}{3}-\policy^{2}_3(x^2_3,x^3_3)\right)\\
    \notag+&(\lambda^1_{1,(1,3)})^\top\left(u^1_3-\policy^{1}_3(x^1_3,x^2_3)\right)\\
    +&(\lambda^1_{1,(1,2)})^\top\left(u^1_2-\policy^{1}_2(x^1_2,x^2_2)\right).\label{eq: l11}
\end{align}
We can now list the KKT conditions for agent 1's problems. Besides primal feasibility, at a Nash equilibrium we must have:
\begin{align}
\begin{aligned}
&\nabla_{u^1_3,x^1_4,x^2_4,x^3_4}\mathcal{L}^1_3=0,\\
&\nabla_{u^1_2,u^1_3,\ufbindex{3}{3},x^1_3,x^2_3,x^3_3,x^1_4,x^2_4,x^3_4}\mathcal{L}^1_2=0,~\text{and}\\
&\nabla_{u^1_1,u^1_2,u^1_3,\ufbindex{3}{3},\ufbindex{2}{3},\ufbindex{3}{2},x^1_2,x^2_2,x^3_2,x^1_3,x^2_3,x^3_3,x^1_4,x^2_4,x^3_4}\mathcal{L}^1_1=0\\
\end{aligned}
\label{eq: kkt of example game}
\end{align}
Using \Cref{eq: l13,eq: l12,eq: l11,eq: kkt of example game} and primal feasibility, one can find the values of the Lagrange multipliers as functions of agent states and controls. Plugging them into \Cref{eq: kkt of example game} and repeating the procedure for agents 2 and 3 yields Riccati-like equations. In this game, given the cyclical nature of the interleaved information, corresponding equations for other agents can be produced by changing agent indices $1\rightarrow2$, $2\rightarrow3$, and $3\rightarrow1$. 

To this end, analyzing the stationarity conditions backwards in time allows us to find Lagrange multiplier values.
For example, $\lambda^1_{2,(1,3)}$ can be shown to be 0, because
\begin{align*}
    \nabla_{u^1_3}L^1_2&=2R^{11}_3u^1_3-[B^1_3]^T\eta^1_{2,(1,4)}+\lambda^1_{2,(1,3)}=0,\\
    \nabla_{u^1_3}\mathcal{L}^1_3&=2R^{11}_3u^1_3-[B^1_3]^T\eta^1_{3,(1,4)}=0,~\text{and}\\
    u^1_3&=\policy^{1}_3(x^1_3,x^2_3)
\end{align*}
yield $\eta^1_{3,(1,4)}=\eta^1_{2,(1,4)}$, and thus $\lambda^1_{2,(1,3)}=0$. 

Similarly, one can verify that $\lambda^i_{t,(i,m)}=0~ \forall ~t\in[m],~ \forall~ m \in [3],~\forall ~i\in[3]$. It can also be shown that not all multipliers values are needed in order for the equilibrium controls and states to be found. For example, the value of $\uindex{1}{2}$ can be found through the stationary conditions $\nabla_{\uindex{1}{2},\xindex{1}{3}}\mathcal{L}^1_2=0$ without needing to find $\gendynlagplayer{1}{3}{2}{3}$. Furthermore, conditions $\nabla_{\xindex{i}{t}}\mathcal{L}^j_k=0$ and $\nabla_{\xindex{i}{t}}\mathcal{L}^j_m=0$ yield $\eta^j_{k,(i,t)}=\eta^j_{m,(i,t)} \forall ~k,m \in \{ 1,\dots,t^*\}$, where $t^*$ is the first time that finding the control requires the use of the corresponding Lagrange multiplier. These results are incorporated back into \Cref{eq: kkt of example game}, which, along with primal feasibility, allows the derivation of Ricatti-like equations.

\addtolength{\textheight}{-12cm}   %

\section{Conclusion}
Realistic multi-agent scenarios often exhibit interleaved information structures, where agents observe only a subset of other agents at each decision-making timestep. In contrast, existing dynamic game literature primarily focuses on canonical open-loop and feedback information structures, which assume that agents observe either only the initial state or the full state of all agents at every timestep. Motivated by this gap, we present two main contributions. First, we develop a systematic procedure to represent dynamic games with interleaved information as Mathematical Program Networks (MPNs). Second, for linear-quadratic (LQ) games, we leverage the MPN formulation to derive Riccati-like equations that characterize Nash equilibria. We illustrate our approach through a three-agent LQ game with a cyclic information structure.
Our framework provides a foundation for analyzing more general classes of dynamic games, and future work should investigate interleaved information games beyond the LQ setting and scenarios where the interleaved information structure is not known a priori and evolves with the agents’ decisions.

{
\bibliographystyle{IEEEtran}
\bibliography{IEEEabrv,references}
}

\end{document}